# A SIMPLE DEFORMATION OF SPECIAL RELATIVITY

CLEMENS HEUSON
*Zugspitzstr. 4*
*D-87493 Lauben, Germany*
*clemens.heuson@freenet.de*

A deformation of special relativity based on a dispersion relation with an energy independent speed of light and a symmetry between positive and negative energy states is proposed. The deformed Lorentztransformations, generators and algebra are derived and some consequences are discussed.

## 1. Introduction

In the past few years there was an increased interest in the old idea see Ref. 1, that a fundamental length $\ell$ might exist and play a crucial role in high energy physics. Such a fundamental length could lead to a generalized uncertainty relation and to a deformed dispersion relation. In Ref. 2 the role of a minimum length in quantum gravity research and its connection to a generalized uncertainty relation was discussed. In Ref. 3 quantum deformations of the Poincare algebra with a fundamental mass parameter $\kappa \sim 1/\ell$ were investigated. More recently in Ref. 4 the minimum length was introduced as kinematical property in so called doubly special relativity theories (DSR), involving two invariant scales, a velocity scale and a length scale or energy/momentum scale. In Ref. 5 a second example of DSR theory was proposed. Finally in Ref. 6 a whole class of DSR theories was considered and a possible connection to Snyder's noncommutative geometry in Ref. 1 was shown. The existence of an invariant minimal length is in clear contradiction to standard Lorentz-Fitzgerald contraction, so one would expect that special relativity must be changed somehow.

In most papers the fundamental length is identified with the Planck length $l_P = \sqrt{\hbar G / c^3} \approx 1.6 \cdot 10^{-35} m$. Usually Newton's constant *G* determining the strength of gravity is considered as a fundamental constant. However due to quantum field theory one would expect, that *G* is a running coupling like the couplings of electromagnetic, weak and strong interactions and therefore Planck's length cannot be a fundamental constant. One is therefore temptated to introduce a fundamental length $\ell$, interpreted as observer independent minimum observable distance or wavelength of particles [4]. Nevertheless on dimensional grounds there must exist a relation $\ell = \rho \cdot l_P$ with $\rho$ a numerical constant presumably of order one. If $\ell$ is very small, then the question, why is gravity so weak, is "reduced" to the question, why is the fundamental length so small. The fundamental constant $\ell$ constitutes the basis of a deformation of special relativity, which can be obtained in the limit $\ell \to 0$.

In the DSR theories proposed so far, there is an asymmetry between positive and negative energy states. In this letter we propose an DSR theory with an energy independent speed of light avoiding this asymmetry and discuss the deformed Lorentztransformations, generators and algebra.

## 2. Deformed dispersion relation

Consider the Magueijo Smolin model in Ref. 5 based on the dispersion relation $(E^2 - p^2)/(1-\ell E)^2 = \mu_0^2$. As noted there the symmetry of positive and negative energy values is broken, therefore one may wish to seek for a symmetric solution. It seems easy to remedy the asymmetry by replacing the denominator with $1-\ell^2 E^2$. A general deformed spatial isotropic dispersion relation with a symmetry between positive and negative energy states and an energy independent speed of light can be written as $F^2 p^2 = \mu_0^2$, where $F = F(\ell^2 E^2)$. Requiring further $F(E=1/\ell) = \infty$ the simplest expression for a deformed dispersion relation is given by

$$\frac{E^2 - p^2}{1-\ell^2 E^2} = \mu_0^2 \qquad (1)$$

Here $\mu_0$ denotes the invariant Casimir mass to be distinguished from the rest mass $m_0$ of the particle. The relation with a plus sign in the denominator can be obtained by the replacement $\ell \to i\ell$. An important consequence of (1) is that the region $|E| > 1/\ell$ is excluded, since the left side would become negative if $E^2 - p^2 > 0$. Furthermore as easily seen from (1) the speed of light is independent of energy as in the model in Ref. 5. The connection with the rest mass is deduced from above by putting $E = m_0$ and $p = 0$.

$$\mu_0 = \frac{m_0}{\sqrt{1-\ell^2 m_0^2}} \qquad (2)$$

Since $\mu_0$ and $\ell$ are invariants the rest mass $m_0$ also must be one. Inserting (2) in (1) and solving for $m_0^2$ results in a dispersion relation equivalent to (1)

$$\frac{E^2 - p^2}{1-\ell^2 p^2} = m_0^2 \qquad (3)$$

In general if $\mu_0^2$ is an invariant, then a function $g(\mu_0^2)$ is also invariant. Therefore (1) and (3) must be invariant under the same deformed Lorentztransformations and are equivalent in this sense, as also will be shown in section 3 and 4. Another example is provided by $\mu^2 = \mu_0^2/(1+(\ell^2/2)\mu_0^2)$ yielding a dispersion relation with an Euclidean deformation $(E^2 - p^2)/(1-(\ell^2/2)(E^2 + p^2)) = \mu^2$. In special relativity the Casimir invariant equals the squared rest mass, and one may take this condition to select dispersion relation (3). We note that (1) and (3) are incompatible to the definition of the particle mass as $1/\mu = \lim_{p \to 0} 1/p \cdot \partial E/\partial p$ with $p = |p|$ postulated for example in Ref. 7 for the Magueijo Smolin model.

Equation (3) can be rewritten as $E^2 = p^2(1-\ell^2 m_0^2) + m_0^2$, which also may be obtained by another reasoning. Consider the standard dispersion relation $p^\mu p_\mu = m_0^2$ and generalize it to a hypothetical 5-dimensional space $p^a p_a = m_0^2$ (a= 0..4), avoiding any speculation about the physical meaning of the extra dimension. By putting $p^4 p_4 = \ell^2 m_0^2 \boldsymbol{p}^2$ one just arrives at the above relation.

## 3. Deformed Lorentztransformations

As observed in Ref. 8 a deformed dispersion relation is sufficient to determine the nonlinear Lorentztransformations. We denote by $\pi^\mu = (\pi^0, \boldsymbol{\pi})$ the linear 4-momentum of special relativity and by $p^\mu = (p^0, \boldsymbol{p})$ the physical high energy 4-momentum obeying the transformations

$$\pi = Fp \ , \ p = F^{-1}\pi \qquad (4)$$

with the nonlinear function $F(p)$ and its inverse $F^{-1}(\pi)$. Compared to the notation in Ref. 8 we have interchanged $F$ and $F^{-1}$, the reason for this is that it is mostly $F$ which appears in our equations. Taking as metric $diag(\eta) = (-1,+1,+1,+1)$ the Casimir of special relativity $C = -\pi^2$ transforms into $C = -F^2 p^2$. $\pi$ transforms according the standard linear Lorentztransformation $\pi' = \Lambda\pi$, from which together with (4) the nonlinear deformed Lorenztransformation $L$ can be obtained as

$$p' = Lp = F^{-1}\Lambda F p \qquad (5)$$

For example the deformed Lorentztransformations of the Magueijo Smolin model can be derived via this procedure. For dispersion relation (3) the equations (4) read

$$\pi = Fp = \frac{p}{\sqrt{1-\ell^2 \boldsymbol{p}^2}} \ , \ p = F^{-1}\pi = \frac{\pi}{\sqrt{1+\ell^2 \boldsymbol{\pi}^2}} \qquad (6)$$

Now from (5) and (6) the deformed Lorentztransformations $L$ in the 1 direction are derived as

$$\begin{aligned} p'^0 &= A\gamma(p^0 - vp^1) \\ p'^1 &= A\gamma(p^1 - vp^0) \\ p'^2 &= Ap^2 \\ p'^3 &= Ap^3 \end{aligned} \qquad (7)$$

$$A = (1 - \ell^2(p^0)^2 + \gamma^2\ell^2(p^0 - vp^1)^2)^{-1/2}$$

where $\gamma = (1-v^2)^{-1/2}$. From (5) and (6) $A$ is at first obtained in the form $A = (1 - \ell^2(p^1)^2 + \gamma^2\ell^2(p^1 - vp^0)^2)^{-1/2}$, which can be rewritten in the form given in (7) using $\gamma^2 - 1 = v^2\gamma^2$. If we had started with dispersion relation (1) instead of (3) then in (6) the replacements, $\boldsymbol{p}^2 \to (p^0)^2$, $\boldsymbol{\pi}^2 \to (\pi^0)^2$ would directly result in (7).
The transition to units with $\hbar, c \neq 1$ is performed by the replacements $\ell \to \ell/\hbar$, $v \to v/c$ showing that (7) can only be valid for quantum particles with $\hbar \neq 0$ and not for macroscopic bodies. For $\ell \to 0$ or small energy and momentum with $A \to 1$ (7) reduces to the standard Lorenztransformation. In the limit $v \to 0$ one has $p' = p$ as in special relativity, however in the limit $v \to 1$ one finds $|p'^0| = |p'^1| = 1/\ell$, $p'^2 = p'^2 = 0$, contrary to special relativity, where $p'^0, p'^1 \to \infty$, $p'^2 = p^2$, $p'^3 = p^3$. The velocity addition rule from (7) with $v'^1 = p'^1 / p'^0$ is identical to the one in special relativity similar as in Ref. 5, especially for $v^1 = 1$ one receives $v'^1 = 1$ and therefore the invariance of c is respected.
The identification of velocity with $v = p/E$ just like in Ref. 5 is of course problematic as discussed in Ref. 9, since it disagrees with the Hamiltonian definition $v = dE/dp$ coming from the Heisenberg equation $\dot{x}_i = v_i = i[H, x_i]$.

It is easy to proove that the first Casimiroperator in the form (3) or (1) is indeed invariant under the deformed Lorentztransformations. Applying (7) to a massive particle in a rest frame with $p^0 = m_0$ and $\mathbf{p} = 0$ gives $A\gamma = (1 - v^2(1 - \ell^2 m_0^2))^{-1/2}$ and thereby the energy-velocity and momentum-velocity relations satisfying the deformed dispersion relation (1) or (3). Alternatively from the classical expressions $\pi^0 = \gamma m_o$ and $\boldsymbol{\pi} = \gamma m_0 \mathbf{v}$ and from (6b) one also arrives at (8).

$$p^0 = \frac{m_0}{\sqrt{1 - v^2(1 - \ell^2 m_0^2)}} \quad , \quad \mathbf{p} = \frac{m_0 \mathbf{v}}{\sqrt{1 - v^2(1 - \ell^2 m_0^2)}} \qquad (8)$$

For $v = 0$ one has $p^0 = m_0$, $\mathbf{p} = 0$ as in special relativity and for $v = 1$ one has $p^0 = 1/\ell$ and $|\mathbf{p}| = 1/\ell$ yielding a strict cutoff of energy and momentum. We note here that the dispersion relation with $\ell \to i\ell$ would not lead to a cutoff in energy and momentum.

Massless particles like photons obey the relation $p^2/(1 - \ell^2 \mathbf{p}^2) = 0$ and therefore satisfy $p^0 = |p^1|$ for a photon moving in the x direction. For a photon with $p^0 = p^1 = 1/\ell$ moving in the x-direction one obtains $p'^0 = p'^1 = 1/\ell$ so $\ell$ is indeed an invariant under the deformed boosts of (7).

The modified Doppler shift in a nonrelativistic regime obtained from (7) is given by $\Delta p^0 / p^0 = v(1 - \ell^2 (p^0)^2)$. Clearly the deviations from special relativity are more difficult to detect since they depend on $\ell^2$ instead on $\ell$ as in Ref. 5.
As discussed in literature, see for example Ref. 10,11 and references therein, DSR theories may eventually provide a solution of the cosmic ray puzzle, the observation of ultra high energy cosmic rays beyond the GZK cut-off obtained from the interaction with the cosmic microwave background. The corrected threshold formula is given by $E_{th} = F^{-1} E_{th0}$ see Ref. 10, where $E_{th0}$ is the special relativistic value. With $F^{-1} = 1/\sqrt{1 + \ell^2 E_{th0}^2}$ for relation (1) one observes, that the threshold is lowered instead of raised, as occurs for the Magueijo Smolin model, and therefore the present model does not provide a solution to this problem. Replacing $\ell \to i\ell$ would raise the threshold, but the amount is to small to provide a solution to the cosmic ray puzzle. Possible ways out have been discussed in Ref. 10,11.

The nonlinear addition of 4-momenta cf. Ref. 8 could be derived from $\pi_{tot} = \sum \pi_i$ resulting together with (4) in $p_{tot} = F^{-1} \sum F p_i$. As discussed in Ref. 10 within this definition there is the problem that sets of N particles may have energies greater than $1/\ell$ though obeying the same deformed dispersion relation. One possible solution is to replace $F_\ell^{-1} \to F_{\ell/N}^{-1}$ in $p_{tot}$ yielding for our dispersion relation (1) choosed for comparison with Ref. 10

$$p_{tot} = \frac{\sum p^{(i)}(1-\ell^2 p^{0(i)2})^{-1/2}}{\sqrt{1+(\ell/N)^2(\sum p^{0(i)}(1-\ell^2 p^{0(i)2})^{-1/2})^2}} \qquad (9)$$

For particles with identical energies and momenta or $p^{0(i)} \ll 1/\ell$ one has the standard addition rule. Finally by multiplying (3) with the scalar wavefunction $\Phi(p)$ one obtains the deformed Klein Gordon equation in momentum space, while the corresponding deformed Dirac equation is given by

$$(\frac{\gamma^\mu p_\mu}{\sqrt{1-\ell^2 \boldsymbol{p}^2}} - m_0)\Psi(p) = 0 \qquad (10)$$

## 4. Deformed algebra

In this section we derive the infinitesimal generators corresponding to (7). One way is to write down the infinitesimal transformations by introducing the rapidity $\xi$ defined by $\tanh(\xi)=v$ and using $\gamma=\cosh(\xi) \approx 1$, $v=\tanh(\xi) \approx \xi$ and $A \approx 1+\xi\ell^2 E p_x$. For a general transformation here in momentum space $p'^\mu = f^\mu(p^\nu,\xi)$ the infinitesimal generators are defined as, see Ref. 12, $N = -i(\partial f^\mu/\partial\xi)_{\xi=0} \partial/\partial p^\mu$ yielding the deformed boost generator.

There exists however another way to obtain the deformed generators generally for an DSR theory with energy independent speed of light defined by equation (4) from the undeformed generators $M_{\mu\nu} = i(\pi_\nu \frac{\partial}{\partial \pi_\mu} - \pi_\mu \frac{\partial}{\partial \pi_\nu})$. The momenta of special relativity transform according $\pi = Fp$, the transformation of derivatives can be expressed via the chain rule together with (4b)

$$\frac{\partial}{\partial \pi^\mu} = \frac{\partial p^\nu}{\partial \pi^\mu}\frac{\partial}{\partial p^\nu} = (F^{-1}\delta_\mu^\nu + \frac{\partial F^{-1}}{\partial \pi^\mu}\pi^\nu)\frac{\partial}{\partial p^\nu} \qquad (11)$$

Applying this derivative to the function $F^{-1}(\pi)$, setting on the right side of (11) $F^{-1}(\pi) = 1/F(p)$ obtained from (4a) and (4b), and solving for $\frac{\partial F^{-1}}{\partial \pi^\mu}$ one finds

$$\frac{\partial F^{-1}}{\partial \pi^\mu} = -\frac{b_\mu}{F^2}, \quad b_\mu = \frac{F_\mu}{F+p^\nu F_\nu} \qquad (12)$$

where $F_\mu = \partial F/\partial p^\mu$. Inserting (12) back into (11) we finally get for the derivatives expressed through nonlinear momenta $p_\mu$

$$\frac{\partial}{\partial \pi^\mu} = \frac{1}{F}(\frac{\partial}{\partial p^\mu} - b_\mu p^\nu \frac{\partial}{\partial p^\nu}) \qquad (13)$$

Putting (13) together with (4a) in the undeformed generators we get for the deformed generators

$$M_{\mu\nu} = i(p_\nu \frac{\partial}{\partial p_\mu} - p_\mu \frac{\partial}{\partial p_\nu} + B_{\mu\nu} p^\lambda \frac{\partial}{\partial p^\lambda}) \qquad (14)$$

where $B_{\mu\nu} = p_\mu b_\nu - p_\nu b_\mu$. The dilatation generator $D = p^\lambda \frac{\partial}{\partial p^\lambda}$ in the deformed boosts first observed in Ref. 5 appears quite general in DSR theories with energy independent speed of light. For spatially isotropic deformed dispersion relations, where $F$ is a function of $E$ (or $\boldsymbol{p}^2$), one obtains $B_{ij} = 0$ and therefore only the boost generators are deformed. One can easily check that these deformed generators leave invariant the deformed Casimir $F^2 p^2$ but not $p^2$. The commutator of the generators with momenta is obtained as

$$[M_{\mu\nu}, p_\sigma] = i(\eta_{\mu\sigma} p_\nu - \eta_{\nu\sigma} p_\mu + B_{\mu\nu} p_\sigma) \qquad (15)$$

The commutators $[M_{\mu\nu}, M_{\rho\sigma}]$ and thereby the Lorentz and the rotation sector of the algebra remain undeformed as can be seen from the expression involving $\pi_\mu$. Denoting $p^0 = -p_0 = E$, $p^i = p_i = (p_x, p_y, p_z)$ we have for the Magueijo Smolin model $1/F = 1 - \ell E$, $b_\mu = \ell \delta_{\mu 0}$, $B_{ij} = 0$, $B_{oi} = -\ell p_i$ yielding the deformed generators derived in Ref. 5. For our model with dispersion relation in the form (1) we have $1/F = \sqrt{1 - \ell^2 E^2}$, $b_\mu = \ell^2 E \delta_{\mu 0}$ or in the form (3) $1/F = \sqrt{1 - \ell^2 \boldsymbol{p}^2}$, $b_\mu = \ell^2 p_i \delta_{\mu i}$ (no summation over i), and $B_{ij} = 0$, $B_{oi} = -\ell^2 E p_i$ in both cases. Thereby the deformed Lorentz generators are

$$N_i = M_{0i} = i(p_i \frac{\partial}{\partial E} + E \frac{\partial}{\partial p_i} - \ell^2 E p_i p^\lambda \frac{\partial}{\partial p^\lambda}) \qquad (16)$$

The commutators of deformed generators and momenta are obtained as

$$\begin{aligned}[N_i, E] &= i p_i (1 - \ell^2 E^2) \\ [N_i, p_j] &= i E (\delta_{ij} - \ell^2 p_i p_j) \end{aligned} \qquad (17)$$

The deformed Lorenzgenerators and thereby the deformed algebra appear to be equivalent for dispersion relations (1) and (3) as was stated in section 2.

## 5. Summary

In summary we have proposed a simple model for a deformation of special relativity with an invariant length scale based on the dispersion relation in the form (1) or (3) respecting particle antiparticle symmetry together with an constant speed of light independent of energy. The model should be testable since it yields a strict cutoff of energy and momentum together with deformed energy-, momentum- velocity relations. We also derived the deformed Lorenztransformations, the infinitesimal generators and the deformed algebra for DSR models with an energy independent speed of light.

Starting with an deformed dispersion relation and following the procedure in sections 2-4 one can investigate a whole class of DSR theories. In principle one also can consider models with varying speed of light, defined by the dispersion relation $F_1^2 E^2 - F_2^2 \boldsymbol{p}^2 = \mu_0^2$, see Ref. 10,13. A varying speed of light is for example obtained as $\hat{c} = E/|\boldsymbol{p}| = F_2/F_1$, which could provide an alternative explanation of the horizon problem in cosmology.

Important open problems of all DSR theories are the description of macroscopic bodies and the corresponding spacetime sector, see Ref. 14 for a recent review. Nevertheless theories with modified Lorentztransformations remain an attractive alternative for ultra high energy physics, which may be testable in a near future.

**Note added:** The present paper is based on a preprint with the same title from 5.8.2002 cited in Ref. 14. In the mean time the same dispersion relation was proposed by Rama in hep-th/0209129 and more recently by Kimberley, Magueijo, Medeiros in gr-qc/0303067.